\documentclass[5p, number, sort&compress]{elsarticle}
\usepackage{import}
\usepackage[dvipsnames, table]{xcolor}
\usepackage{amsmath, graphicx}
\usepackage{subcaption}
\usepackage{url}
\usepackage{mathrsfs}
\usepackage{colortbl}
\usepackage{hyperref}
\usepackage{tikz}
\usepackage{adjustbox}
\usepackage{mathtools,nccmath}
\usetikzlibrary{calc,positioning}
\usepackage{nomencl}
\makenomenclature
\usepackage{rotating}
\makeatletter
\providecommand{\doi}[1]{%
  \begingroup
    \let\bibinfo\@secondoftwo
    \urlstyle{rm}%
    \href{http://dx.doi.org/#1}{%
      doi:\discretionary{}{}{}%
      \nolinkurl{#1}%
    }%
  \endgroup
}
\makeatother
\usepackage{bookmark}
\usepackage[nameinlink,noabbrev,capitalize]{cleveref}

\renewcommand{\cref}{\Cref}

\begin{document}

\title{Gabor-Type Holography Solved Analytically for Complex-valued Phase Disks}
\author[1]{Jesper Gl\"uckstad\corref{cor1}}
\ead{jegl@mci.sdu.dk}
\author[1]{Andreas Erik Gejl Madsen}
 \ead{gejl@mci.sdu.dk}

\cortext[cor1]{Corresponding author}
\address[1]{SDU Centre for Photonics Engineering \\ University of Southern Denmark \\ DK-5230 Odense M, Denmark}
\begin{abstract}
Solving the holography equation has long been a numerical task. While effective, the numeric approach has its own set of limitations. Relying solely on numerical approaches often obscures the intricate interplay and influence of the individual terms within the equation. This not only hampers a deeper understanding of the underlying physics but also makes it challenging to predict or control specific outcomes.
In this study, we address these challenges by leveraging our recently published \cite{gluckstad_new_2023} updated Fraunhofer diffraction expression. This approach allows us to derive an analytic solution for complex-valued phase disks in on-axis holography. This solution facilitates the direct computation of each term's influence within the holographic equation, paving the way for a more profound comprehension and application of the holographic process. When compared to experimental results and the numeric Fresnel diffraction solution, our analytic approach shows impressive accuracy, considering the inherent approximations. Notably, it remains precise for Fresnel numbers that extend well beyond the traditionally accepted boundaries of the Fraunhofer regime.
\end{abstract}


\maketitle

\section{Introduction}
\noindent
Holography \cite{gabor_new_1948} has, since its invention, found uses in numerous areas of academia and industry, including microscopy \cite{pacheco_adaptive_2022,ren_automatic_2019,brault_automatic_2022,oconnor_deep_2020,descoteaux_efficient_2017,huang_holographic_2021,denis_inline_2009,guo_lensfree_2022,ma_quantitative_2021,castaneda_video-rate_2021,madsen_-axis_2023}, light-shaping \cite{gerchberg_practical_1972,madsen_algorithmic_2021,madsen_efficient_2023,banas_holo-gpc_2017,madsen_holotile_2022,banas_light_2017,pang_speckle-reduced_2019,liu_symmetrical_2006,wu_adaptive_2021}, particle trapping and manipulation \cite{curtis_dynamic_2002,grier_holographic_2006,burnham_holographic_2006,agarwal_manipulation_2005,suarez_optical_2021,sun_theory_2008}, cryptography \cite{li_efficient_2019,rajabalipanah_real-time_2020,guo_stokes_2022,yang_visual-cryptographic_2018}, etc.

Interestingly, while holography's foundational principles were laid by Denis Gabor, he faced the persistent challenge of the twin-image problem. At the time, there was no analytical solution to quantify or eliminate this twin-image issue, leading Gabor to effectively abandon his on-axis holography research after exploring various optical setups. This decision stands in stark contrast to the subsequent proliferation and significance of holography in diverse fields. The ability to regain both amplitude and phase information of an optical wavefront after its recording by an intensity-sensitive detector is a key factor of its popularity. The holographic equation, which reveals this property of holographic image capture, can be written as the intensity of the superposition of two incident wavefronts - the object and reference:
\begin{align}
	\mathcal{I} &= (\mathcal{O} + \mathcal{R})(\mathcal{O} + \mathcal{R})^\ast \nonumber \\
	&= \mathcal{R}\mathcal{R}^\ast + \mathcal{O}\mathcal{O}^\ast + \mathcal{O}\mathcal{R}^\ast + \mathcal{O}^\ast\mathcal{R} \label{eq:holo-gabor}
\end{align}
where $\mathcal{I}$ is the resultant intensity, $\mathcal{O}$ and $\mathcal{R}$ are the object and reference wavefronts, respectively, and $^\ast$ denotes the complex conjugate. The four terms comprising the equation are $\mathcal{R}\mathcal{R}^\ast$, the directly transmitted unscattered reference wave, $\mathcal{O}\mathcal{O}^\ast$, the scattered object wave interfering with itself, thus denoted here as a ``self-inteference" term, and $\mathcal{O}\mathcal{R}^\ast$ and $\mathcal{O}^\ast\mathcal{R}$, real and virtual images of the original object wavefront, scaled by the reference wave, containing both amplitude and phase information of the object. Typically, due to the complex conjugate, the term $\mathcal{O}^\ast\mathcal{R}$ is often denoted as the ``twin-image" of the object.

Now, to extract the amplitude and phase of the original object wavefront, the term $\mathcal{O}\mathcal{R}^\ast$ must be isolated to the greatest extent. In off-axis holography, in which the illuminating light is split into an object beam and an angled reference beam, thus introducing a carrier wave, the four terms of \cref{eq:holo-gabor} can ``simply" be separated in Fourier space \cite{leith_reconstructed_1962, leith_wavefront_1963,leith_wavefront_1964}. However, when discussing on-axis holography, the object illuminating light source serves as both the object and reference wave. It is conceptually divided into the two waves, but originates from the same light beam. Thus, both wavefronts propagate axially in-line, not allowing for spatial separation of the terms.
\begin{figure}
	\centering
	\includegraphics[width=\columnwidth]{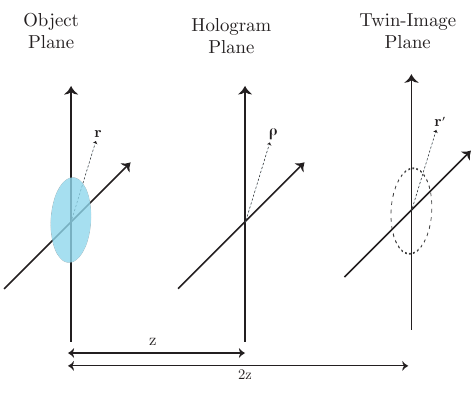}
	\caption{Illustration of the different planes and their coordinate notations as used in the derivations. The object plane contains the phase disks at coordinates $\mathbf{r}$. The hologram plane is denoted by $\boldsymbol{\rho}$, and the twin-image plane by $\mathbf{r^\prime}$.}
	\label{fig:notation}
\end{figure}

The reconstruction of the object beam in on-axis holography is typically undertaken with iterative numerical methods, namely derivations of the well-known Gerchberg-Saxton algorithm \cite{gerchberg_practical_1972,wu_adaptive_2021,pang_speckle-reduced_2019,chang_speckle-suppressed_2015,liu_symmetrical_2006,yuan_digital_2021}. While good results can be achieved, the algorithm is inherently non-convex, not ensuring convergence to the global minimum. In addition, the iterative nature limits the reconstruction speed. In the last decade, machine learning methods \cite{sinha_lensless_2017,ren_end--end_2019,rivenson_phase_2018,ju_learning-based_2022,huang_holographic_2021,madsen_-axis_2023} tackling the same problem of wavefront retrieval have been introduced showing promising results. While inference can be performed at incredible reconstruction speed, the training of the networks require both long training time and many thousand training samples.

In this paper, we present an approach to describe the holography equation in \cref{eq:holo-gabor} analytically for disk-shaped particles in a uniform medium. Using our recently published addendum to the Fraunhofer propagation method \cite{gluckstad_new_2023}, the analytical description maintains good accuracy for Fresnel numbers well above the scope of the original Fraunhofer propagation domain (usually $N_f < 1/8$). Such an analytic description allows for in-depth understanding of each term in the equation, and, in turn a simple method for quantifying the influence of both the self-interference and twin-image term.
\begin{figure}
	\centering
	\includegraphics[width=\columnwidth]{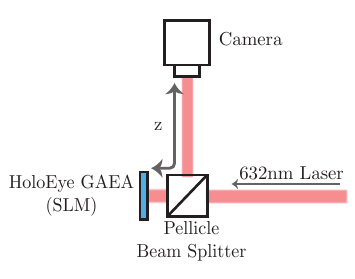}
	\caption{Illustration of the optical setup used for validating the analytic solution. A camera on a translation stage records holograms at varying propagation distances. The phase disks are synthesized by a HoloEye GAEA SLM, illuminated by a collimated laser $632.8$ nm source.}
	\label{fig:optical-setup}
\end{figure}
\section{Analytic Holographic Reconstruction}
In this analysis, the reference light at the image capturing sensor is given by an amplitude-normalized plane wave;
\begin{equation}
	\mathcal{R} = 1
\end{equation}
Considering a sparse sample of disk-shaped micro-particles diluted in a medium, the object light can, to a first approximation, be expressed as the free-space propagated light from the transmission through the disks, each of which expressed as
\begin{equation}
	\textrm{disk}(\mathbf{r}, \mathbf{r}_0) = A_d e^{(-i\varphi_d)} \textrm{circ}\left(\frac{|\mathbf{r} - \mathbf{r}_0|}{\Delta r} \right)
\end{equation}
where $\mathbf{r}=(x, y)$ is the coordinate vector, $\mathbf{r}_0$ is the disk center, $\Delta r$ is the radius of the disk, $A_d$ is the disk amplitude transmission and $\varphi_d$ is the relative phase retardation of the disk.
The propagated disk-transmitted light, ``carved" out from the reference light $\mathcal{R}$, can be written as a convolution with the paraxial Fresnel phase kernel
\begin{equation}
	\frac{e^{ikz}}{i\lambda z}e^{\frac{ik}{2z} r^2} \Rightarrow e^{iqr^2}
\end{equation}
Where $r = \sqrt{x^2 + y^2}$, $q = \frac{k}{2z} = \frac{\pi}{\lambda z}$, and where the multiplicative constants have been omitted. Thus, object wave $\mathcal{O}$ for an axis-centered disk incident on the image capturing sensor at a distance $z$, is
\begin{equation}
	\mathcal{O} = \left[ A_d e^{-i\varphi_d} - \mathcal{R} \right] \textrm{circ}\left(\frac{r}{\Delta r} \right) \otimes e^{iqr^2}
\end{equation}
Thus, substituting into \cref{eq:holo-gabor}, the four holographic terms on the image sensor are given by
\begin{align}
	\mathcal{I} =& (\mathcal{O} + \mathcal{R})(\mathcal{O}+\mathcal{R})^\ast \nonumber \\
	=& \left( \left[ A_d e^{-i\varphi_d} - \mathcal{R} \right] \textrm{circ}\left(\frac{r}{\Delta r} \right) \otimes e^{iqr^2} + \mathcal{R} \right) \nonumber \\
	& \cdot\left( \left[ A_d e^{i\varphi_d} - \mathcal{R} \right] \textrm{circ}\left(\frac{r}{\Delta r} \right) \otimes e^{-iqr^2} + \mathcal{R} \right) \nonumber \\
	=& \mathcal{R}\mathcal{R}^\ast + |\left[ A_d e^{-i\varphi_d} - \mathcal{R} \right] \textrm{circ}\left(\frac{r}{\Delta r} \right) \otimes e^{iqr^2}|^2 \nonumber \\
	&+ \left[ A_d e^{i\varphi_d} - \mathcal{R} \right] \textrm{circ}\left(\frac{r}{\Delta r} \right) \otimes e^{-iqr^2} \nonumber \\
	&+ \left[ A_d e^{-i\varphi_d} - \mathcal{R} \right] \textrm{circ}\left(\frac{r}{\Delta r} \right) \otimes e^{iqr^2} \label{eq:hologram-exp}
\end{align}
Now, these captured hologram terms can be backpropagated from the image sensor to the original object location, $-z$, by convolution with the complex conjugate Fresnel kernel $e^{-iqr^2}$:
\begin{align}
	C =& \;1\nonumber \\
	&+ |\left[ A_d e^{-i\varphi_d} - 1 \right] \textrm{circ}\left(\frac{r}{\Delta r} \right)\otimes e^{iqr^2}|^2 \otimes e^{-iqr^2} \nonumber \\
	& + \left[ A_d e^{i\varphi_d} - 1 \right] \textrm{circ}\left(\frac{r}{\Delta r} \right) \otimes e^{-i2qr^2} \nonumber \\
	& + \left[ A_d e^{-i\varphi_d} - 1 \right] \textrm{circ}\left(\frac{r}{\Delta r} \right) \label{eq:holo-reconstruct}
\end{align}
Where the unscattered reference light is normalized to $\mathcal{R}\mathcal{R}^\ast = \mathcal{R}=1$. 
\cref{eq:holo-reconstruct} shows \cref{eq:holo-gabor} for the specific case of the micro-disks. The equation contains the same four terms; the background $\mathcal{R}$, the self-interference term, $\mathcal{O}\mathcal{O}^\ast$, which now a defocused quadratic term that is often ignored in the literature as it is considered negligible compared to the others, the twin-image $\mathcal{O}^\ast \mathcal{R}$, a twice-defocused virtual image of the object located $2z$ behind the object, and the desired original object term $\mathcal{O}$.

\subsection{Analytic Form of the Twin-Image Term}
The twin-image term of \cref{eq:holo-reconstruct},
\begin{equation}
	C_\textrm{twin} = \left[ A_d e^{i\varphi_d} - 1 \right] \textrm{circ}\left(\frac{r}{\Delta r} \right) \otimes e^{-i2qr^2}, \label{eq:twin-term}
\end{equation}
can be easily calculated by numeric integration using the Fresnel diffraction integral for circular apertures \cite[p. 102]{goodman_introduction_2017}. However, as we are aiming to arrive at a completely analytic expression for calculation speed and increased understanding, we look toward the Fraunhofer regime and its ease of calculation.
Typically, a Fresnel number in the area of $N_f < 1/8$ is required for physically accurate propagation simulation using the Fraunhofer expression. In our previous publication \cite{gluckstad_new_2023}, we showed how the small argument approximation, $\left(1 - e^{i \pi N_f}\right) \approx -i\pi N_f$, that is commonly applied in the derivation of the Fraunhofer propagation expression rapidly becomes invalid as the Fresnel number nears its imposed limit, and that, by simple substitution of the full $\left(1 - e^{i \pi N_f}\right)$, the accuracy of the Fraunhofer expression can be maintained up to at least twice the ordinary limit ($N_f < 1/4$).

Thus, with the updated Fraunhofer propagation expression \cite{gluckstad_new_2023}, the twin-image term in \cref{eq:twin-term} can be written as
\begin{align}
	C_\textrm{twin} =& \left[ A_d e^{i\varphi_d} - 1 \right] \left(1 - e^{i \pi N_f / 2}\right) e^{i \pi \frac{r^2}{2\lambda z}} \nonumber \\
	& \cdot\left[\frac{2J_1 (2\pi r \Delta r/(2\lambda z))}{2\pi r\Delta r / (2\lambda z)} \right]
\end{align}
where the Fresnel number is halved to account for the twice-defocused propagation.

\subsection{Analytic Form of the Self-Interference Term}
Using the same updated Fraunhofer expression, the self-interference term of \cref{eq:holo-reconstruct} can be rewritten as
\begin{align}
	C_\textrm{self} =& \left|\left[ A_d e^{-i\varphi_d} - 1 \right] \left(1 - e^{i \pi N_f}\right) e^{i \pi \frac{\rho^2}{\lambda z}}  \right. \nonumber \\
	&\cdot\left. \left[\frac{2J_1 (2\pi \rho \Delta r /(\lambda z)}{2\pi \rho \Delta r  / (\lambda z)} \right]\right| ^2 \otimes e^{-iqr^2} \label{eq:self-pre}
\end{align}
where $\rho$ is the radial coordinate in the hologram plane.
In order to reformulate this expression as analytic, we make an important approximation. It is commonly known that an Airy disk can be approximated by a Gaussian function \cite{engelbrecht_resolution_2006,zhang_gaussian_2007}. As such, we can simplify \cref{eq:self-pre} by separating the squared absolute and substituting the Airy disk for an appropriate Gaussian beam profile:
\begin{align}
	C_\textrm{self} =& \left|\left[ A_d e^{-i\varphi_d} - 1 \right] \left(1 - e^{i \pi N_f}\right)\right| ^2  \nonumber \\
	&\cdot\left| \left[\frac{2J_1 (2\pi \rho \Delta r /(\lambda z)}{2\pi \rho\Delta r  / (\lambda z)} \right]\right| ^2 \otimes e^{-iq\rho^2} \nonumber \\
	=& \left|\left[ A_d e^{-i\varphi_d} - 1 \right] \left(1 - e^{i \pi N_f}\right)\right| ^2 e^{\frac{-\rho^2}{w_0^2}}\otimes e^{-iq\rho^2}
\end{align}
Where $w_0$ is the Gaussian beam waist parameter. Upon closer inspection of the convolution, we can directly obtain \cite{goodman_introduction_2017}
\begin{equation}
	e^{\frac{-\rho^2}{w_0^2}}\otimes e^{-iqr^2} = \frac{w_0}{w_z}e^{\frac{\rho^2}{w_z^2}} e^{\frac{-i\pi\rho^2}{\lambda R_z}} e^{i\psi_z}
\end{equation}
Where $w_z$ is the beam width along the propagation axis, $R_z$ is the wavefront curvature, and $\psi_z$ is the Gouy phase. These parameters can be directly calculated from the parameters of the Airy disk the Gaussian function approximates. The full-width at half-maximum (FWHM) of an Airy function is given by $\textrm{FWHM}_\textrm{Airy} =z \cdot \sin{\left(1.025 \frac{\lambda}{2\Delta r}\right)}$. The beam waist parameter of a Gaussian beam with identical FWHM can be calculated as
\begin{equation}
	w_0 = \frac{\sqrt{2\ln(2)}}{2} \textrm{FWHM}_\textrm{Airy}
\end{equation}
From the beam waist parameter, the remaining parameters are calculated as follows
\begin{itemize}
	\item $z_0 = \frac{\pi w_0^2}{\lambda}$
	\item $w_z = w_0 \sqrt{1 + \left(\frac{z}{z_0}\right)^2}$
	\item $R_z = z \left[ 1 + \left( \frac{z_0}{z}\right)^2\right]$
	\item $\psi_z = \tan^{-1}\frac{z}{z_0}$
\end{itemize}
Thus, the self-interference can be approximated as
\begin{equation}
	C_\textrm{self} = \left|\left[ A_d e^{-i\varphi_d} - 1 \right] \left(1 - e^{i \pi N_f}\right) \right| ^2 \frac{w_0}{w_z}e^{\frac{\rho^2}{w_z^2}} e^{\frac{-i\pi\rho^2}{\lambda R_z}} e^{i\psi_z}
\end{equation}

\section{Experimental Work}

With the analytic versions of the twin and self-interference terms, the full reconstruction in \cref{eq:holo-reconstruct} becomes
\begin{figure*}[ht!]
	\centering
	\includegraphics[width=\textwidth]{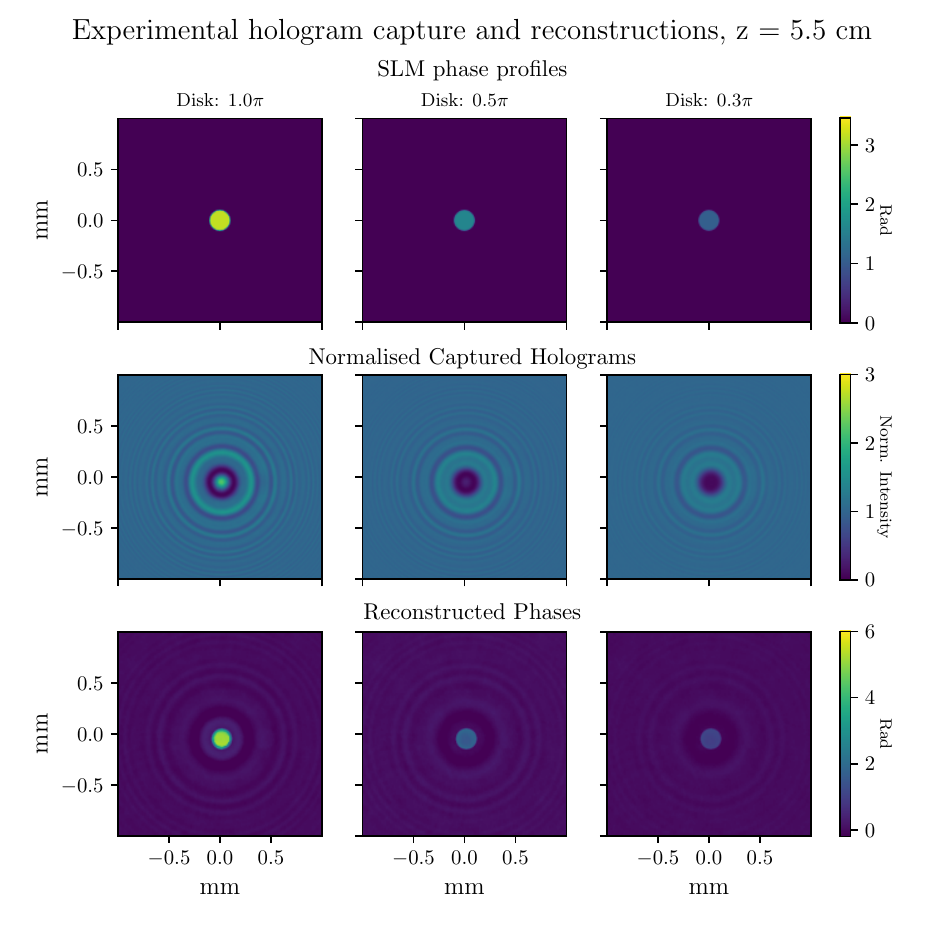}
	\caption{Phase profiles as displayed on the SLM, captured resulting holograms, and the reconstructed phase profiles for the $100\mu$m as captured experimentally at a propagation distance of $z=5.5$ cm at varying disk phases.}
	\label{fig:experimental-slm}
\end{figure*}

\begin{align}
	&C_a(r) = 1 \nonumber \\
	& + \left|\left[ A_d e^{-i\varphi_d} - 1 \right] \left(1 - e^{i \pi N_f}\right) \right| ^2 \frac{w_0}{w_z}e^{\frac{\rho^2}{w_z^2}} e^{\frac{-i\pi\rho^2}{\lambda R_z}} e^{i\psi_z} \nonumber \\
	& + \left[ A_d e^{i\varphi_d} - 1 \right] \left(1 - e^{i \pi N_f / 2}\right) e^{i \pi \frac{r^2}{2\lambda z}}
	\left[\frac{2J_1 (2\pi r \Delta r/(2\lambda z))}{2\pi r\Delta r / (2\lambda z)} \right] \nonumber \\
	& + \left[ A_d e^{-i\varphi_d} - 1 \right] \textrm{circ}\left(\frac{r}{\Delta r} \right) 
\end{align}

To validate the accuracy of this fully analytic expression, we compare it to both experimental data and the numeric solution to \cref{eq:holo-reconstruct} using the Fresnel diffraction integral.

\begin{figure*}
	\centering
	\includegraphics{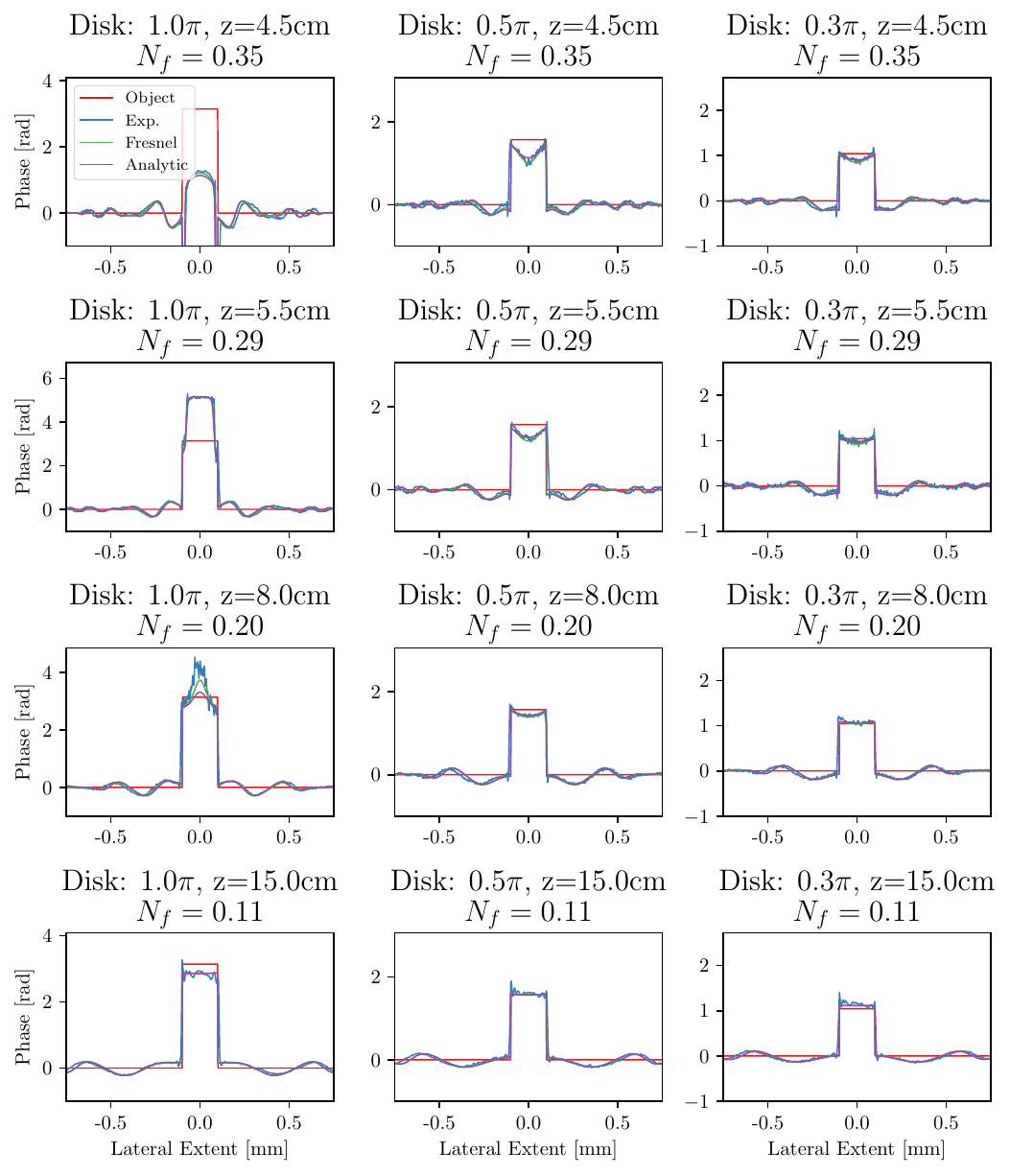}
	\caption{Comparison between novel analytical solutions (purple), numeric Fresnel diffraction solutions (green), and experimentally captured and numerically reconstructions (blue) for phase disks of varying phase shifts and at increasing propagation distances.}
	\label{fig:comp-fres-anal}
\end{figure*}

\begin{figure*}
	\centering
	\includegraphics{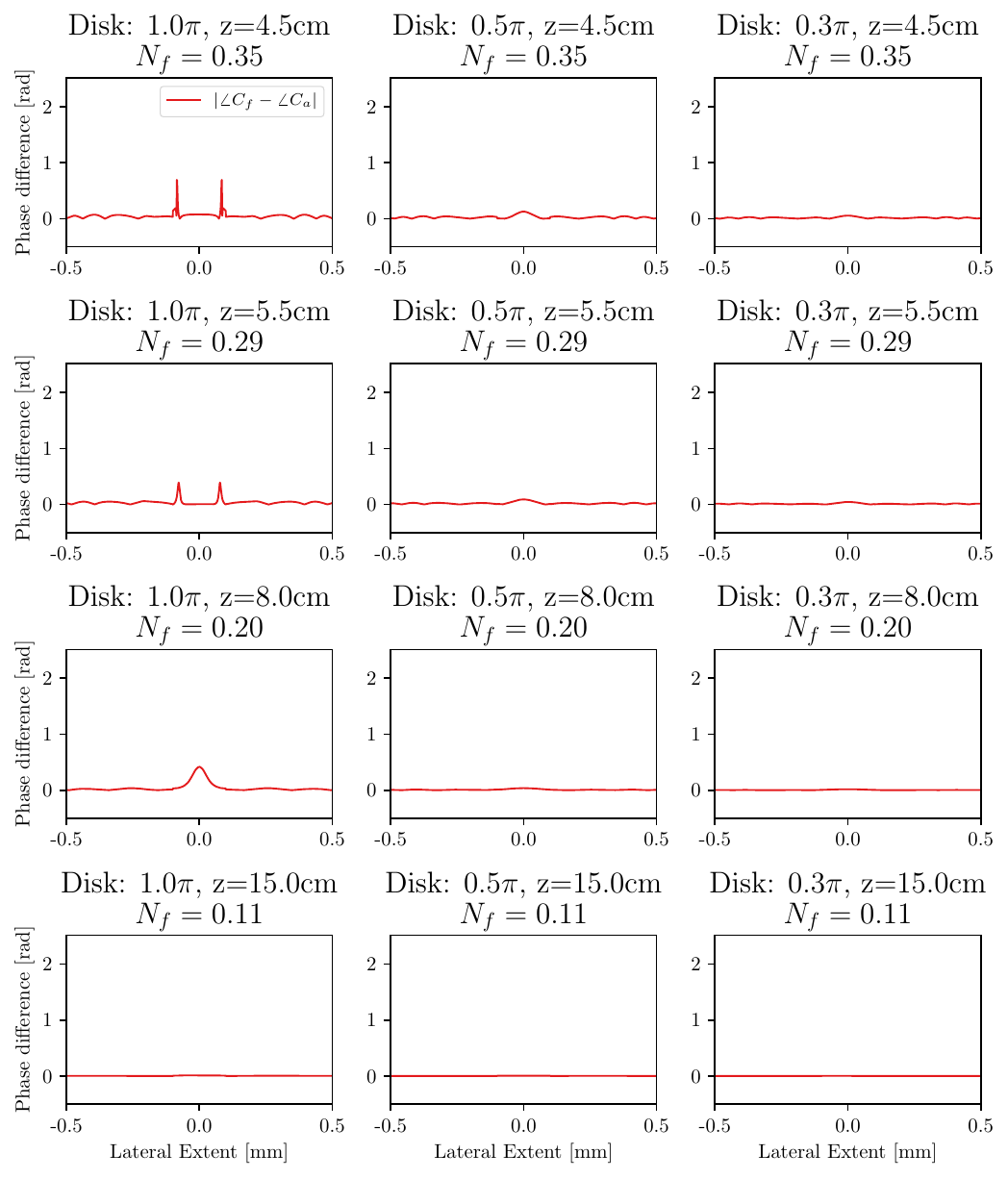}
	\caption{The absolute difference in object phase between the analytic and Fresnel solutions for disks with varying phase shifts at increasing propagation distances.}
	\label{fig:phase-difference}
\end{figure*}

An illustration of the experimental setup is seen in \cref{fig:optical-setup}. Using a HoloEye GAEA spatial light modulator (SLM), incoming expanded and collimated laser light from a Thorlabs Helium-Neon 632.8 nm laser is phase-modulated. In the center of the SLM we display phase patterns of a single disk with a radius of $100 \mu$m and incremental phase values of $\pi$, $\pi/2$, and $\pi/3$, respectively. Then, for each disk, a camera is translated axially to capture the disk holograms at various propagation distances. Each disk hologram is numerically backpropagated to the SLM plane by way of the angular spectrum method (ASM)\footnote{Fine-tuning to find the correct propagation distance is necessary, taking into account discrepancies between the measured and actual propagation distance.}. The shown phase profiles, captured holograms, and reconstructed phase profiles are visualized in \cref{fig:experimental-slm}. Finally, a cross section through the centre of the reconstructed phase disk is extracted, such that it can more easily be compared to the numeric and analytic methods.

The analytic solution is also compared to the numeric solution of \cref{eq:holo-reconstruct}, which, when implementing the Fresnel diffraction integral for a circular disk, can be expressed as 
\begin{align}
	&C_\textrm{f}(r) = 1 \nonumber \\
	& + \left|\left[ A_d e^{-i\varphi_d} - 1 \right] 2\pi  \int\limits_0^{\sqrt{N_f}} r \,e^{i\pi \frac{r^2}{\lambda z}}J_0\left(\frac{2\pi \rho r}{\lambda z}\right) dr
 \right|^2 \otimes e^{-iq\rho} \nonumber \\
	& + \left[ A_d e^{i\varphi_d} - 1 \right] \frac{2\pi}{i} e^{i \pi \rho^{\prime 2}} \int\limits_0^{\sqrt{N_f/2}} r^\prime \,e^{i\pi \frac{r^{\prime 2}}{2\lambda z}}J_0\left(\frac{2\pi r r^\prime}{2\lambda z} \right) dr^\prime \nonumber \\
	& + \left[ A_d e^{-i\varphi_d} - 1 \right] \textrm{circ}\left(\frac{r}{\Delta r} \right) 
\end{align}
Where the integrals are evaluated numerically and the backpropagation in the self-interference term is performed via ASM. Both the numeric and analytic reconstructions are calculated for each measured propagation distance from the physical experiment.

\section{Results}
The reconstructed phases for each method are shown in \cref{fig:comp-fres-anal}. As can be seen, both the numeric Fresnel expression and the new analytic expression model the experimental data extremely well. Even the abrupt change in the experimentally reconstructed phase for the $\pi$-disk at $z=4.5 \textrm{ cm}$ is captured in the analytic reconstruction, as well as the ``beating" of the waves surrounding the reconstructed object. While the introduction of the approximations that allow for our analytic solution inevitably introduces errors as compared to the numerical solution, which is shown in \cref{fig:phase-difference}, the error is practically negligible, especially when comparing to experimental data with inherent noise.
Note must be taken here to appreciate the Fresnel numbers that are at work here. The \textit{fully} analytic expression for the holographic reconstruction is maintaining a similar reconstruction accuracy as the more correct numeric solution for Fresnel numbers at least as high as $N_f = 0.35$, more than twice that of the established requirement for valid Fraunhofer diffraction.

\section{Discussion}
It is important to realize that the primary objective of deriving an analytic expression and subsequent reconstruction isn't necessarily to retrieve the precise phase profile of the object in question. Instead, it allows us to unveil the inherent implications and constraints posed by the holographic equation.

Consider, for example, experimental outcomes like those observed for the $\pi$-disk at positions $z=4.5$ cm and $z=5.5$ cm. Given the significant discrepancies between the experimentally reconstructed phase and the actual object phase, one might be inclined to believe that some form of error has found its way into either the experimental process or the numeric reconstruction.

However, it is clear that most of the reconstruction errors are directly explainable by the analytic solution. This understanding is pivotal, as it shifts the perspective from viewing these as simple errors to recognizing them as inherent characteristics of the holographic process.

Furthermore, the analytic expression offers a valuable toolset. For instance, it is now trivial to directly calculate the unwanted influence of both the twin-image and the self-interference term, or, inversely, to solve for a required propagation distance or disk radius to obtain a certain specified attenuation of these interfering terms.
We here provide several example usecases where the analytic expressions may be used as design guidelines. First, consider a sample consisting of disk-like transparent objects of varying sizes. The minimal propagation distance can be calculated required to obtain clear reconstructions:
\begin{itemize}
	\item Identify the largest disk size in the sample
	\item Determine acceptable $N_f$ for the desired twin-image attenuation
	\item Calculate the corresponding required minimal propagation distance between sample and image sensor
\end{itemize}

Alternatively, in a different scenario, the location of a disk object can be determined in a single-parameter optimization, given that the dimensions and phase shift of the disk object is known:
\begin{itemize}
	\item Identify the size and phase shift of the disk
	\item Capture a hologram with the disk at an unknown axial distance
	\item Apply \cref{eq:hologram-exp} analytically by utilizing the updated Fraunhofer expression \cite{gluckstad_new_2023}, and fit the result to the captured hologram with $r$ and $z$ as the fitting parameters
	\item Optionally, determine if the propagation distance provides sufficent twin-image suppression
\end{itemize}
The latter example can of course be applied to several more cases by swapping the knowns and unknowns of the system e.g., the disk phase can be identified if the disk radius and propagation distance are known, and the disk radius can be found for known disk phase and propagation distance.

\section{Conclusion}
\label{sec:conclusion}
Utilizing our updated Fraunhofer diffraction expression \cite{gluckstad_new_2023} and a Gaussian approximation of the Airy disk, we have arrived at an analytic solution to Gabor holography for complex-valued phase disks.xx The solution allows for the direct calculation of the influence of each term in the holographic equation, opening it up for much deeper understanding and utilization. The analytic solution is compared to both experimental results and the numeric Fresnel diffraction solution, and it is shown that it maintains good accuracy for Fresnel numbers well outside the scope of the conventionally established Fraunhofer regime.

\section*{Acknowledgements}
\noindent
This work has been supported by the Novo Nordisk Foundation, Denmark (Grand Challenge Program; \\ NNF16OC0021948), the Innovation Fund Denmark, and Radiometer Medical ApS.

\newpage
\bibliographystyle{elsarticle-num-names}
\bibliography{references}

\appendix
\gdef\thesection{\Alph{section}}
\makeatletter
\renewcommand\@seccntformat[1]{\appendixname\ \csname the#1\endcsname.\hspace{0.5em}}
\makeatother

\end{document}